\documentclass[twocolumn]{aastex62}

\graphicspath{{./}{figures/}}
\usepackage{natbib}
\submitjournal{ApJ}

\shorttitle{Radiation Feedback in Dwarf Galaxies}
\shortauthors{Emerick et. al.}

\begin{document}

\title{Stellar Radiation is Critical for Regulating Star Formation and Driving Outflows in Low Mass Dwarf Galaxies}

\correspondingauthor{Andrew Emerick}
\email{emerick@astro.columbia.edu}

\author{Andrew Emerick}
\affil{Department of Astronomy, Columbia University, New York, NY, 10027, USA}
\affil{Department of Astrophysics, American Museum of Natural History, New York, NY 10024}

\author{Greg L. Bryan}
\affiliation{Department of Astronomy, Columbia University, New York, NY, 10027, USA}
\affiliation{Center for Computational Astrophysics, Flatiron Institute, 162 5th Ave, New York, NY, 10003, U.S.A}

\author{Mordecai-Mark Mac Low}
\affil{Department of Astrophysics, American Museum of Natural History, New York, NY 10024}
\affiliation{Department of Astronomy, Columbia University, New York, NY, 10027, USA}

\nocollaboration

% Letter word count:
%    Text (including captions) : 3500
%    Abstract                  : 500
%    # Figures                 : 5
%       max panels per fig     : 9
%    # references              : 50

\begin{abstract}
%
%
% NOTE: Overleaf includes abstract words in word count in text... subtract
%       this from the total.
%
%
Effective stellar feedback is used in models of galaxy formation to 
%regulate star formation and 
drive realistic galaxy evolution. Models typically include energy injection from supernovae as the dominant form of stellar feedback, often in some form of sub-grid recipe. However, it has been recently suggested that pre-SN feedback (stellar winds or radiation) is necessary in high-resolution simulations of galaxy evolution to properly regulate star formation and properties of the interstellar medium (ISM). Following these processes is computationally challenging, 
%Following stellar winds and ionizing radiation is computationally challenging, 
so many prescriptions model this feedback approximately, accounting for the local destruction of dense gas clouds around newly formed stars in lieu of a full radiative transfer calculation. In this work we examine high resolution simulations (1.8~pc) of an isolated dwarf galaxy with detailed stellar feedback tracked on a star-by-star basis. By following stellar ionizing radiation with an adaptive ray-tracing radiative transfer method, we test its importance in regulating star formation and driving outflows in this galaxy. We find that including ionizing radiation reduces the star formation rate (SFR) by over a factor of 5, and is necessary to produce the ISM conditions needed for supernovae to drive significant outflows. We find that a localized approximation for radiation feedback is sufficient to regulate the SFR on short timescales, but does not allow significant outflows. Short and long range radiation effects are both important in driving the evolution of our low-metallicity, low-mass dwarf galaxy. Generalizing these results to more massive galaxies would be a valuable avenue of future research.
\end{abstract}


\section{Introduction} \label{sec:intro}
Historically, simulations of galaxy formation have suffered from the ``overcooling'' problem, whereby the action of self-gravity and radiative cooling alone produces galaxies with far too many stars. This problem has been addressed by employing various models of strong stellar feedback physics which are capable of generating self-regulating star formation in galaxies (see \cite{SomervilleDave2015} and \cite{NaabOstriker2017} for recent reviews). Energy injection from supernovae (SNe) has historically been used as the sole form of feedback. However, this is generally done heuristically as many simulations lack the ability to resolve the Sedov phase of individual SNe. But even with this strong feedback, recent work has argued for the need for pre-SN feedback, from stellar winds and/or stellar radiation \citep[e.g.][]{Hu2016,Hopkins2018}, though this is typically modeled simply as additional energy injection around newly formed stars. The need for additional feedback is confirmed by simulations that are capable of fully resolving individual SN \citep[e.g.][]{Peters2017,Smith2018a,Smith2018b,Hu2018}. However, modeling these processes in detail is challenging, and their competing effects on galactic evolution are poorly constrained.

Radiation from massive stars dominates the total feedback energy output of a stellar population \citep[e.g.][]{Abbott1982,Leitherer1999,Agertz2013}, surpassing the energy ejection of supernova ($\sim 10^{51}$~erg) by two orders of magnitude. If radiation couples effectively to the interstellar medium (ISM), it can be a substantial source of additional feedback. Simulations including stellar radiation feedback followed through radiative transfer or radiation-hydrodynamics schemes have found it to be effective in regulating star formation and driving galactic winds \citep[e.g.][]{WiseAbel2012,Kim2013a,Sales2014,Oshea2015,Rosdahl2015,Pawlik2015,Ocvirk2016,Peters2017}. This occurs in four ways: 1) heating of diffuse gas and preventing the formation of cold, dense star formation regions, 2) destruction of cold, dense gas around recently formed stars, preventing further star formation, 3) momentum input by direct absorption of UV radiation by gas and (in some cases) dust through re-emission and scattering in the infrared, and 4) lowering the typical ISM densities in which SNe occur and greatly increasing their effectiveness. 

However, most works that employ stellar radiation feedback to account for these effects do so using various forms of sub-grid, approximate models to avoid the substantial additional cost of full radiative transfer. Many works use a Str{\"o}mgren approximation whereby the particles / cells within the Str{\"o}mgren radius around a radiating star are heated and kept ionized, with additional approximations made to account for overlapping ionized regions \citep[e.g.][]{HQM2011,Hu2016,Hu2017}. Other works employ some form of energy or momentum injection localized to the region immediately around a star particle \citep[e.g.][]{Agertz2013,Roskar2014,Ceverino2014,Forbes2016}. Although some of these approximate methods account for long-range effects of diffuse radiation \citep{HQM2012,Hopkins2018} most cases treat local radiation {\it only}, confined to energy or momentum injection in a limited physical region around a star particle. This is done both because the local destruction of dense clumps of gas around newly formed stars is commonly believed to be the dominant impact of stellar radiation feedback and because it is computationally less expensive to implement. The role of long-range stellar radiation, once ionizing photons break out of star forming regions, is not well characterized. Indeed it remains to be seen if modeling only the short-range effects of stellar radiation feedback is sufficient.

In this work we use a detailed model for stellar feedback presented in \cite{Emerick2018} (hereafter Paper I) to study the role of radiation feedback in dwarf galaxy evolution. For the first time in a galaxy-scale simulation, we resolve individual HII regions using an adaptive ray tracing radiative transfer method to follow the ionizing radiation from particles that represent individual stars. We focus on addressing two questions: 1) what role does radiation feedback play in regulating star formation, and 2) are the long-range effects of radiation feedback important, or is the local destruction of dense gas the dominant effect.

To investigate these questions, we compare three simulations of the evolution of an isolated, low mass dwarf galaxy. Our fiducial model, containing full stellar radiation feedback, is compared against a run without ionizing radiation feedback, and a run with ionizing feedback limited to the local region around a given star. We discuss our methods in Section~\ref{sec:methods} and present our results ion Section~\ref{sec:results}.

\section{Methods and Initial Conditions} \label{sec:methods}
We refer the reader to Paper I for a more detailed description of our methods. We use the adaptive mesh refinement hydrodynamics code \textsc{Enzo} \citep{Enzo2014} to evolve an idealized, isolated low mass dwarf galaxy. The galaxy is initialized as a smooth exponential disk set in hydrostatic equilibrium with a static dark matter potential \citep{Burkert1995} with $M_{\rm gas} = 2 \times 10^6~$~M$_{\odot}$, $Z_{\rm gas} = 4\times 10^{-4}$, radial and vertical gas scale heights of 250~pc and 100~pc respectively, and $M_{\rm vir} = 2.48 \times 10^9$~M$_{\odot}$, on a grid with a maximum physical resolution of 1.8~pc. We include no initial stellar population, but do include random driving from supernovae as an initial source of feedback up until the onset of star formation.

We include a UV background, tabulated metal line cooling, and a 9 species non-equilibrium chemistry solver using \textsc{Grackle} \citep{GrackleMethod}. Star formation occurs stochastically in cold, dense regions ($n > 200$~cm$^{-3}$, $T < 200$~K) by sampling a \cite{Salpeter1955} IMF from 1~M$_{\odot}$ to 100~M$_{\odot}$ and depositing \textit{individual} star particles over this mass range. We include feedback from stellar winds, AGB winds, FUV and LW band radiation which drives photoelectric heating and H$_2$ dissociation respectively, HI and HeI ionizing radiation, and core collapse and Type Ia SNe using thermal energy injection. FUV and LW band radiation are both taken to be optically thin, with local (cell-by-cell) attenuation alone. Ionizing radiation is followed using radiative transfer, as discussed below. Stellar lifetimes, surface gravity, effective temperature, and radii are set by the initial stellar mass and metallicity through interpolation over the PARSEC \citep{Bressan2012} zero age main sequence values. These properties are used to set the FUV, LW, and ionizing photon fluxes from each star through interpolation on the OSTAR2002 \citep{LanzHubeny2003} grid. We only model the radiation from stars with masses above 8~M$_{\odot}$. 

Our fiducial simulation follows photoionizing radiation using the adaptive ray-tracing radiative transfer method of \cite{WiseAbel2011}. This method places and evolves 48 rays on a HEALPix grid around each emitting source star for both HI and HeI ionizing radiation. Rays are adaptively split as they propagate away from their source to increase the angular resolution such that the solid angle of the ray remains smaller than 1/3 of the cell area. Rays begin on HEALPix refinement level 2, with a maximum refinement level of 13. We include radiation pressure on hydrogen, but do not investigate its importance in this work \citep[see][ and references therein]{Krumholz2018}.

We additionally present a simulation run without any stellar ionizing radiation (``noRT'') and a second simulation that includes ionizing radiation, but deletes all photons that travel more than 20~pc from their source (``shortrad''). This second simulation tests the relative importance of short-range vs. long-range effects of stellar ionizing radiation as a form of feedback. Although this is still more accurate than an approximate method, this is meant to function similarly to methods that include only the local effects of stellar radiation feedback. Each of our three simulations is identical up until the formation of the first star particles.\footnote{We note that the stars from the very first star formation event in each run (21 star particles with a total mass of 114 M$_{\odot}$) are the same. Of these, one particle emits ionizing radiation ($M_* > 8 $~M$_{\odot}$). The sampled stellar masses differ across runs after this point. Stochastic effects from differences in IMF sampling may be important in the regime of low mass, low SFR dwarf galaxies. Multiple re-simulations of identical initial conditions may result in significant scatter in their final properties \citep{Keller2018}. We find our main conclusions are likely to be insensitive to these stochastic variations, as we find that the total radiation output across multiple re-samplings of a fixed star formation history is small after the IMF is fully sampled.}

As shown in Paper I, the maximum densities reached in these simulations is below 10$^{3}$~cm$^{-3}$. Our inability to resolve the high densities in star forming regions ($\sim 10^5$~cm$^{-3}$) means that we likely underestimate initial photon absorption and overestimate the initial long-range effects of newly formed stars. However, in the Milky Way, newly formed O stars have been observed to spend no more than  10 -- 20\% of their main sequence lifetimes embedded in ultracompact HII regions within dense molecular clouds \citep{WoodChurchwell1989}. This short dispersal timescale agrees with high resolution simulations of massive star formation \citep[e.g.][]{Peters2010,Dale2014,Kim2018}. Therefore, we do not consider our neglect of this phase to be likely to lead to substantial errors at the galactic scale.

\section{Results} \label{sec:results}

\begin{figure*}
\centering
\includegraphics[width=0.49\linewidth]{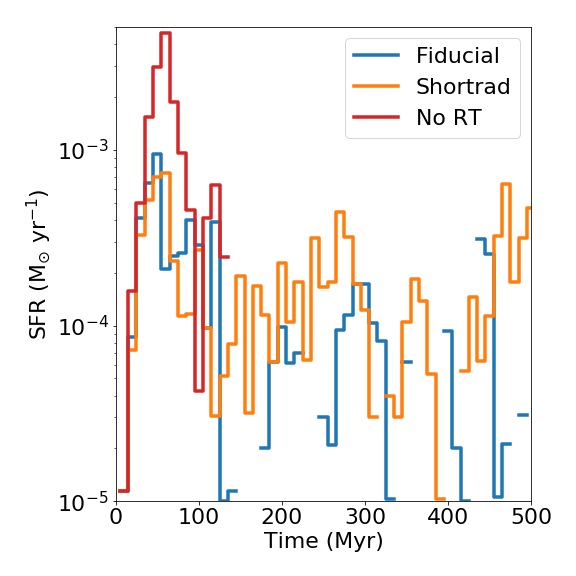}
\includegraphics[width=0.49\linewidth]{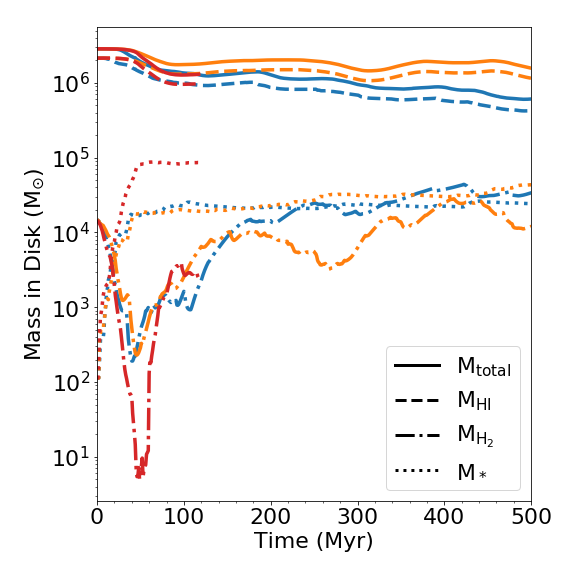}
\caption{The star formation rate (left) and gas and stellar mass (right) of each of our three simulations over time. Time bins with no star formation are left empty.}
\label{fig:sfr_mass_evolution}
\end{figure*}

\begin{figure}
\centering
\includegraphics[width=0.99\linewidth]{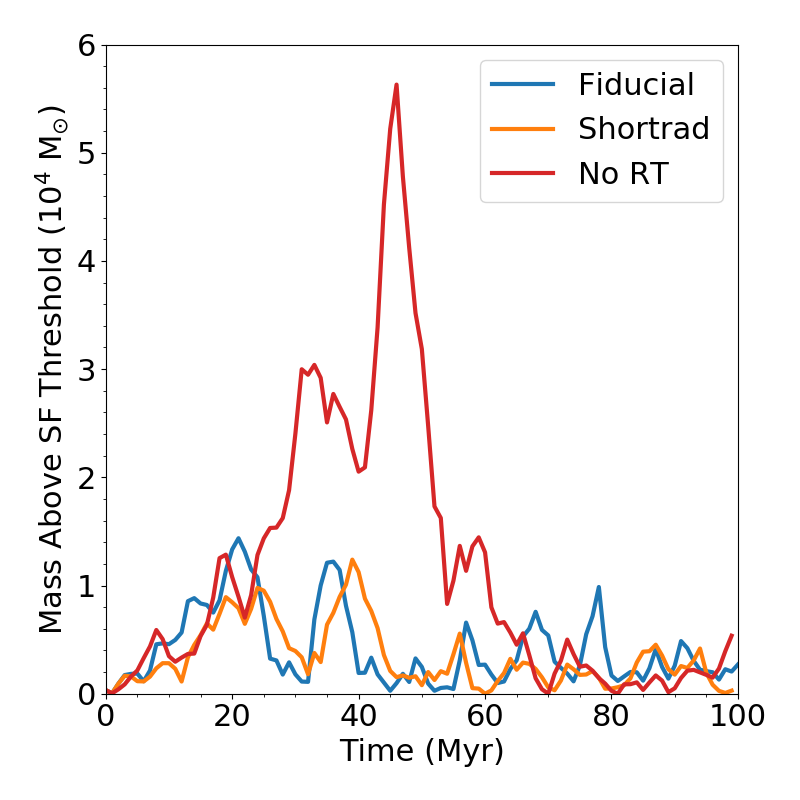}
\caption{The total gas mass at $n > 200$~cm$^{-3}$, our star formation density threshold.}
\label{fig:sf gas}
\end{figure}

We compare the resulting star formation rate (left) and gas mass properties (right) of our three simulations in Figure~\ref{fig:sfr_mass_evolution}. There is a clear, significant contrast between the runs with and without ionizing radiation. Stellar ionizing radiation leads to a factor of $\sim$ 5 reduction in the resulting SFR, as compared to the noRT run. Since the shortrad and fiducial simulations are so similar over the first $\sim$~100~Myr, it is clear that stellar radiation acts to significantly reduce the local star formation efficiency around young, massive stars. Radiation from these stars quickly ionizes and dissipates surrounding dense gas that would otherwise have gone to fuel a significant amount of additional star formation. This is confirmed in Figure~\ref{fig:sf gas}, which shows the gas mass in each simulation above the SF density threshold of $n = 200$~cm$^{-3}$ during the first 100~Myr. While our fiducial and shortrad simulations remain roughly the same here, the noRT run, at its peak, has an order of magnitude more cold, dense gas. \footnote{We are unable to follow the long-term evolution of the noRT galaxy due to computational constraints. Although radiative transfer itself is computationally expensive, this run is substantially more costly due to a lower typical timestep and increased cost in computing the optically thin radiation effects (photoelectric effect and LW dissociation) for the additional star particles.}

Looking again at the first $\sim$100~Myr of simulation time, the effects of ionizing radiation beyond our 20~pc cutoff radius are not significant. However, these two simulations begin to diverge after this point. The shortrad simulation has continual, steady star formation for the entire simulation time, while the star formation rate in our fiducial run is bursty, with periods of active, low-level star formation interspersed with periods of no star formation. The driver of these differences is clear in the right hand panel. Our fiducial run loses a factor of $\sim 2$ more total gas and HI mass (orange and blue lines) as compared to the shortrad simulation. Clearly, galactic winds and outflows are much more effective in with full accounting of stellar ionizing radiation feedback.

\begin{figure*}
\centering
\includegraphics[width=0.32\linewidth]{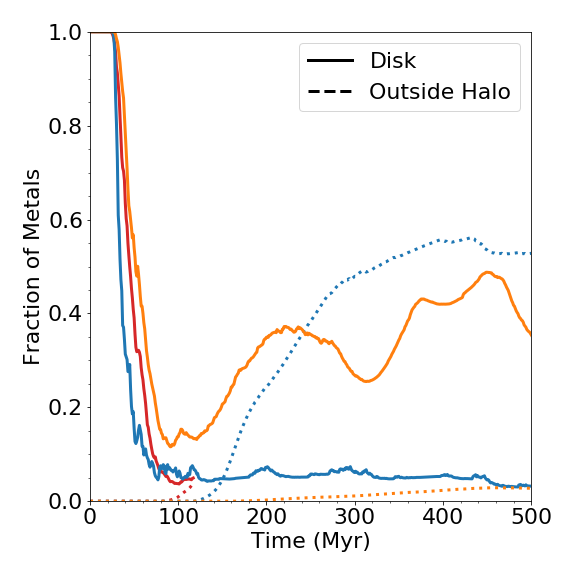}
\includegraphics[width=0.32\linewidth]{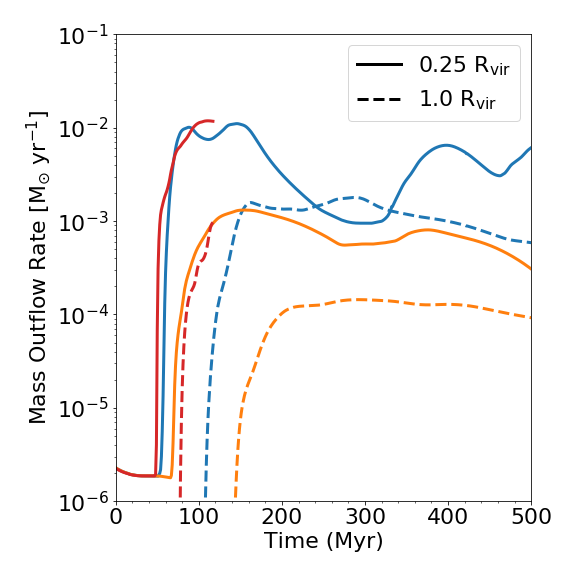} 
\includegraphics[width=0.32\linewidth]{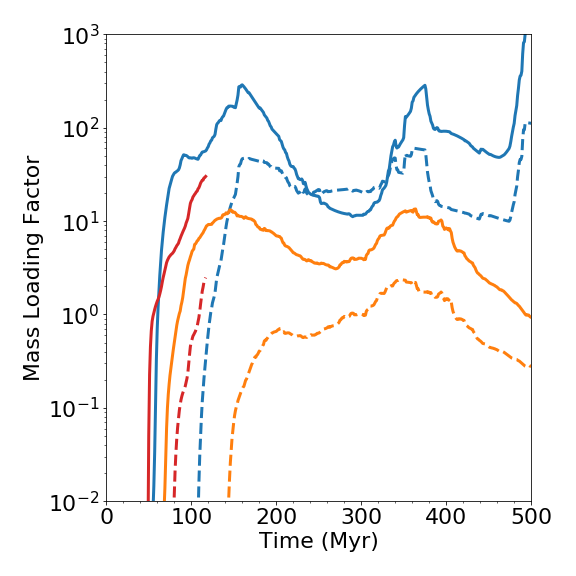}
\caption{{\bf Left:} The fraction of metals contained within the ISM of each galaxy (blue) and the fraction ejected beyond the virial radius (orange). As in Figure~\ref{fig:sfr_mass_evolution}, the fiducial run is given as solid lines, while the noRT and shortrad are dash-dotted and dashed respectively. {\bf Middle:} The mass outflow rate for each run at 0.25~R$_{\rm vir}$ (solid) and R$_{\rm vir}$ (dashed). {\bf Right:} The mass loading factor $\eta$ calculated as the outflow rate divided by the SFR averaged over 100~Myr. Inclusion of diffuse ionization makes an order of magnitude difference in $\eta$ in this case. }
\label{fig:metal_retention}
\end{figure*}

This can be confirmed by examining the metal retention fraction (the fraction of produced metals retained within the disk of the galaxy) and mass outflow rates across simulations. As shown in Figure~\ref{fig:metal_retention}, the mass outflow rate in the fiducial run peaks at an order of magnitude higher at 0.25~R$_{\rm vir}$ than the shortrad simulation, declining only due to a comparative drop off in star formation. The outflow in noRT is only a factor of a few lower than the fiducial run, but it requires a five times higher supernova rate in this simulation to match the same outflow seen with full stellar radiation feedback, so it implies a far lower mass loading factor. Because of this difference in SFR, 
%mm [rewrote to display equation]
the differences across simulations are more significant for our computed mass loading factor 
\begin{equation}
    \eta = \frac{ \dot{M}_{\rm out}}{ <\rm{SFR}>_{100 \rm{Myr}}}.
\end{equation} The brackets indicate time averaging over 100~Myr.
While the fiducial run reaches a peak $\eta$ of a few hundred, shortrad is consistently an order of magnitude or more below this. The noRT simulation is even lower. 

The fiducial run is also the only one of the three with any significant outflow beyond the virial radius. We conclude that radiation feedback allows SNe to be substantially more effective in driving outflows. 

\section{Discussion} \label{sec:discussion}
\begin{figure*}
\centering
\includegraphics[width=0.99\linewidth]{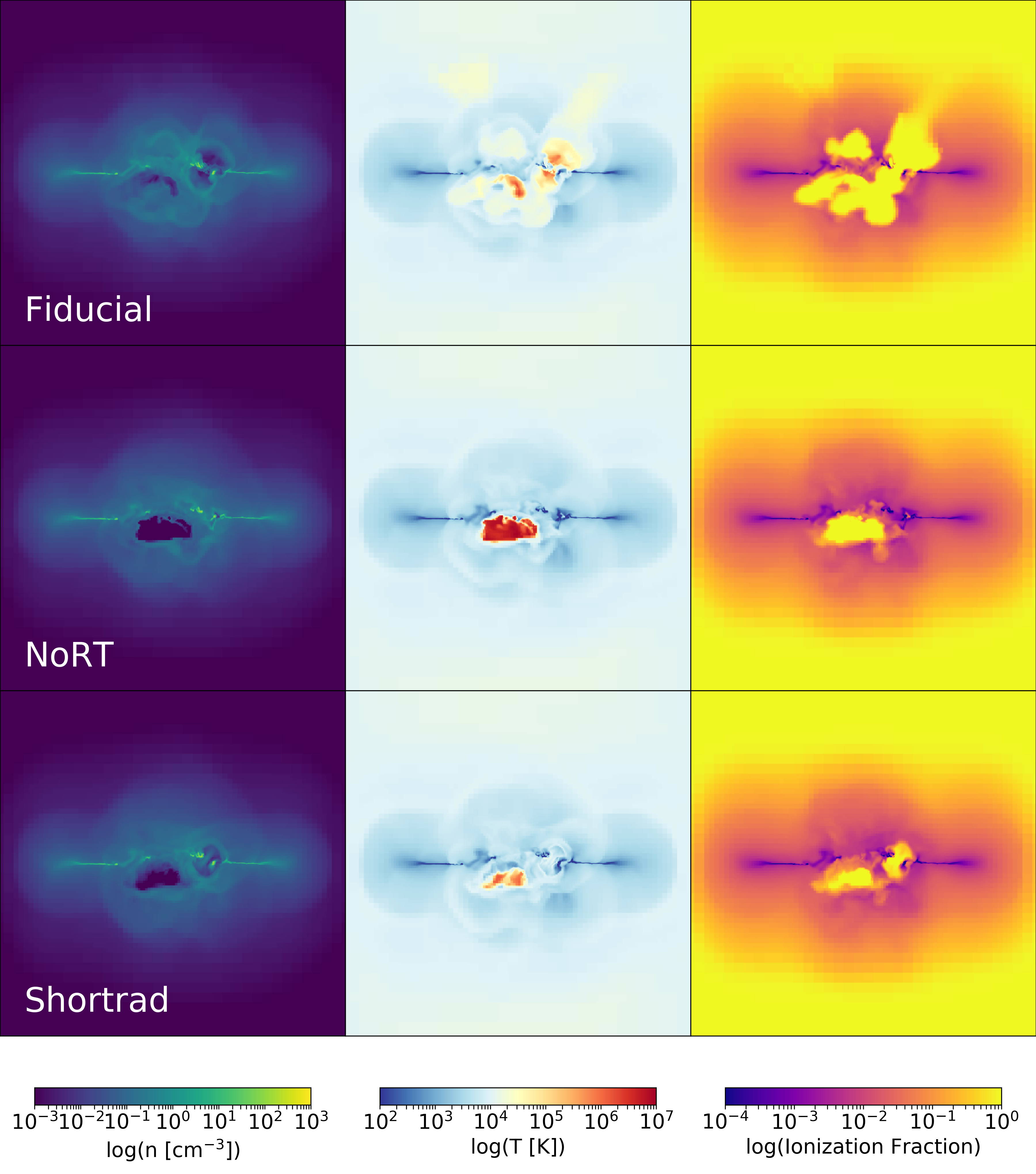}
\caption{Edge-on slices of each simulation showing gas number density (left), temperature (middle), and hydrogen ionization fraction (right) 17~Myr after the formation of the first star in each run. Each panel is 4 kpc x 4 kpc. The full movie of this evolution is included for further clarification of the comparison between this figure and Fig.~\ref{fig:panel2}.}
\label{fig:panel1}
\end{figure*}

\begin{figure*}
\centering
\includegraphics[width=0.99\linewidth]{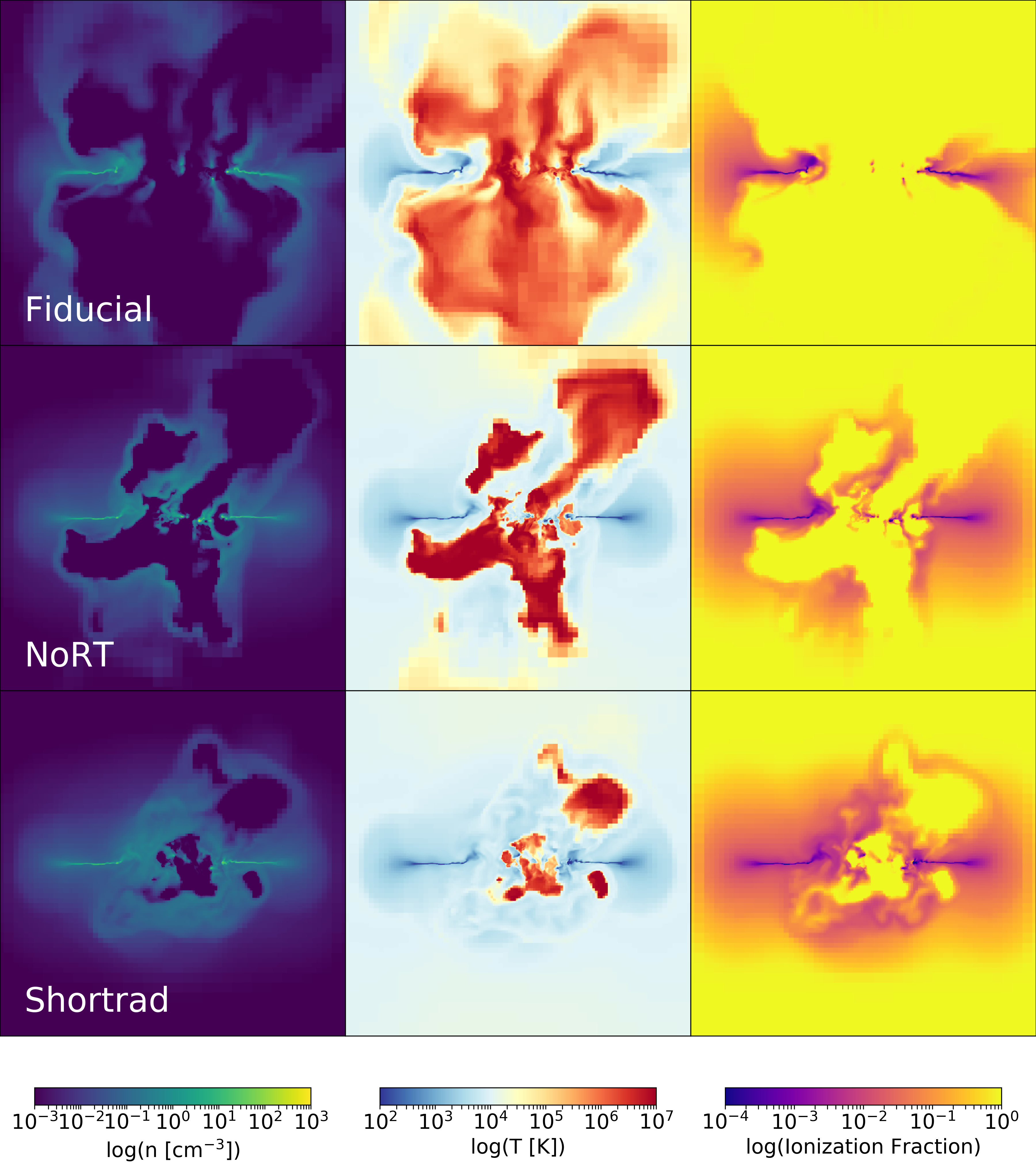}
\caption{Same as Figure~\ref{fig:panel1}, but at 40~Myr.}
\label{fig:panel2}
\end{figure*}

Accounting for feedback from stellar radiation plays a significant role in determining the ability for SN energy to couple to the ISM and therefore drive outflows. We believe this work is novel in examining the
importance of localized ionization vs.\ ionization from a diffuse radiation field far from a single star. Modeling only local stellar ionizing feedback is insufficient to describe the long-term evolution of an isolated dwarf galaxy. To explore the cause of this difference we compare in Figure~\ref{fig:panel1} and~Figure~\ref{fig:panel2} the gas number density (left), temperature (middle), and hydrogen ionization fraction (right) in edge-on slices in each of our simulations at two different times.

Figure~\ref{fig:panel1}, at 17~Myr, compares each simulation just after the first few SNe. Already there are significant differences between the runs. Gas outside the galaxy is warm ($\sim$~10$^{4}$~K) and ionized up to $\sim$500~pc above/below the plane of the disk in our fiducial run. This same gas is cold ($<10^4$~K) and neutral in both other runs. The contrast between the effect of ionizing radiation in the fiducial and shortrad runs is seen most clearly in the ionized region in-plane and to the right of center. This region contains massive stars that are capable of generating enough ionizing radiation to carve a channel out to the halo of our fiducial simulation; this does not occur in the shortrad case. Instead, the HII region is confined by surrounding cold, neutral gas. 

Although the ISM properties within the HII region in each case are quite similar between the simulations, the SNe that eventually go off in this region are confined by the neutral gas in the shortrad simulation, but readily escape through the lower density ionized gas into the galaxy halo in our fiducial case. As these simulations evolve, the existence of these diffuse, ionized channels in the fiducial run easily allow continual and significant outflows from SNe, as shown in  Figure~\ref{fig:panel2}, which shows each simulation 40~Myr after the formation of the first stars. In contrast, the same SNe in the other two simulations are well contained, surrounded by shells of denser, neutral gas. Though they make an important local impact on the ISM, they are unable to drive significant mass loss from the galaxy. In the noRT case, an outflow does eventually develop, but it takes a factor of five increase in SFR, and a corresponding increase in supernova rate, for SNe to finally break out from the neutral gas surrounding the galaxy. 

As shown in Figure~\ref{fig:metal_retention} the differences in ionization structure and its effect on galactic winds has direct consequences for the chemical evolution of our dwarf galaxy. The winds in our fiducial simulation carry nearly all of the metals produced out from the disk of our galaxy. This is the only run in agreement with observations of Local Group dwarf galaxies, both dSph's \citep{Kirby2011} and the gaseous, star forming Local Group dwarf galaxy Leo P \citep{McQuinn2015}, with metal ejection fractions of up to 95\%. This is in stark contrast to the $\sim$30\% retention fraction in our shortrad simulation. This would also influence the chemical enrichment of neighboring galaxies, given the significant differences in metal ejection past the virial radius between these two runs. Clearly the effects of feedback on observable chemical properties of galaxies is a key discriminator among models.

These results show that long-range ionization effects are an important consideration in models of stellar feedback. However,
further study is warranted of how this effect can be approximated without resorting to full radiative transfer calculations. Approximate, Str{\"o}mgren-like feedback models that allow for ionization far from a single source \citep[e.g.][]{Hopkins2018} may be sufficient to capture the effects shown here. Fully localized methods, or methods which set a maximum for the ionization radius may underestimate galactic wind properties in dwarf galaxies. In addition, both methods are mass biased \citep[see the discussion in ][]{Hu2017}, preferentially over-ionizing dense gas that would otherwise be missed by photons leaking through channels carved through lower-density gas. These remaining uncertainties motivate a continued examination of the feedback prescriptions adopted in high resolution simulations of galaxy evolution.

\section{Conclusion}  \label{sec:conclusion}
In agreement with previous work we find that (local) stellar radiation feedback is effective in regulating star formation, but that non-local ionizing radiation is key for driving outflows in our simulations of an isolated, low mass, dwarf galaxy. Simulations run without ionizing radiation feedback have star formation rates a factor of five higher than our fiducial simulation. Despite the lower rate, SNe in our fiducial run are capable of driving larger galactic outflows, aided significantly by the ionizing radiation feedback.

We demonstrate for the first time that simple prescriptions of local stellar radiation feedback fail to reproduce the evolution of our fiducial model. Our simulation with radiation localized to 20~pc around each star particle does effectively regulate star formation on short time scales, predominately by quickly destroying cold, dense gas around young, hot stars. However, this model does not drive the significant outflows seen in our fiducial simulation. Long-range ionizing radiation is important for carving channels allowing the ejection of significant amounts of mass and metals from the SNe. Our simulation with localized radiation feedback retains a significantly higher fraction of metals than expected observationally for low mass dwarf galaxies.  

Finally, we note that we have performed this experiment on only one possible type of galaxy. Its low virial temperature ($\sim10^{4}$~K) makes this galaxy particularly sensitive to the effects of stellar feedback, and ionizing radiation in particular. Examining the role of long-range, diffuse stellar ionizing radiation on star formation and galactic winds in more massive galaxies is an important avenue of future research. 

\acknowledgments
A.E. is funded by the NSF Graduate Research Fellowship DGE 16-44869. G.L.B. acknowledges support from NSF AST-1312888, NASA NNX15AB20G, and NSF AST-1615955. M.-M.M.L. was partly funded by NASA  grant NNX14AP27G and NSF grant AST18-15461. We gratefully recognize computational resources provided by NSF XSEDE through grant number TGMCA99S024, the NASA High-End Computing Program through the NASA Advanced Supercomputing Division at Ames Research Center, Columbia University, and the Flatiron Institute. This work made significant use of many open source software packages, including \textsc{yt}, \textsc{Enzo}, \textsc{Grackle}, \textsc{Python}, \textsc{IPython}, \textsc{NumPy}, \textsc{SciPy}, \textsc{Matplotlib}, \textsc{HDF5}, \textsc{h5py}, \textsc{Astropy}, \textsc{Cloudy} and \textsc{deepdish}. These are products of collaborative effort by many independent developers from numerous institutions around the world. Their commitment to open science has helped make this work possible. 

\software{astropy \citep{astropy} matplotlib \citep{matplotlib}, Numpy \citep{numpy}, SciPy \citep{scipy}, IPython \citep{Ipython}, yt \citep{yt}, Enzo \citep{Enzo2014}, Grackle \citep{GrackleMethod}, CLOUDY \citep{cloudy}, HDF5 \citep{HDF5}}

\bibliography{msbib} 

\end{document}